\documentclass[useAMS,usenatbib]{mn2e}

\title[The extended stellar envelopes of cD galaxies]{Intracluster light and the extended stellar envelopes of cD galaxies: An analytical description}
\author[M.~S. Seigar et al.]{Marc S. Seigar$^{1}$\thanks{E-mail:
mseigar@uci.edu (MSS)}\thanks{McCue Fellow}, Alister W. Graham$^{2}$ and Helmut Jerjen$^{3}$\\
$^{1}$Center for Cosmology, Department of Physics \& Astronomy, University of California, Irvine, 4129 Frederick Reines Hall,\\
Irvine, CA 92697-4575, USA\\
$^{2}$Centre for Astrophysics \& Supercomputing, Swinburne University of Technology, Hawthorn, Victoria 3122, Australia\\ 
$^{3}$Research School of Astronomy \& Astrophysics, Mount Stromlo Observatory, Cotter Road, Weston Creek, ACT 2611, Australia\\
}
\begin{document}

\date{In original form 19 October 2006}

\pagerange{\pageref{firstpage}--\pageref{lastpage}} \pubyear{2007}

\maketitle

\label{firstpage}

\begin{abstract}

We have analysed deep $R$-band images, down to a limiting surface
brightness of 26.5 R-mag arcsec$^{-2}$  (equivalent to 
$\sim$28 B-mag arcsec$^{-2}$),
of 5 cD galaxies to determine the shape of the 
surface brightness profiles of their extended stellar envelopes. 
Both de Vaucouleurs $R^{1/4}$ model and S\'ersic's $R^{1/n}$ model, on
their own, provide a poor description of the surface brightness profiles
of cD galaxies.  This is due to the presence of outer stellar
envelopes, thought to have accumulated over the merger history of the
central cluster galaxy and also from the tidal stripping of galaxies at
larger cluster radii.
We therefore simultaneously
fit two S\'ersic functions to measure the shape
of the inner and outer components of the cD galaxies.
We show that, for 3 out of our 5 galaxies, the surface brightness 
profiles are best fit by an inner S\'ersic model, with indices
$n\sim1-6$, and an outer {\it exponential} component. 
For these systems, 
the galaxy-to-envelope size ratio is 0.1 -- 0.4 and the
contribution of the
stellar envelope to the total $R$-band light (i.e. galaxy + envelope)
is around 60 to 80 per cent (based on extrapolation to a 300 kpc radius).
The exceptions are NGC 6173, for which our surface brightness profile
modelling is consistent with just a single component (i.e. no envelope)
and NGC 4874 which appears to have an envelope with a de Vaucouleurs, rather than 
exponential, profile. 
% We therefore tentatively conclude that there may be no unique surface
% brightness profile which fits the envelopes of cD galaxies.

\end{abstract}

\begin{keywords}
galaxies: elliptical and lenticular, cD -- galaxies: formation -- galaxies: fundamental parameters -- galaxies: haloes -- galaxies: structure
\end{keywords}

\section{Introduction}

First-ranked  galaxies in clusters, also referred to as brightest
cluster galaxies (BCGs), are the brightest and most massive galaxies in the
Universe. They are typically elliptical galaxies (Lauer \& Postman 1992). 
About 20\% of BCGs appear to be 
surrounded by a large, low surface brightness
envelope and are additionally referred to as 
cD galaxies (e.g. Dressler 1984; Oegerle \& Hill 2001). Such cDs reside only
in clusters and groups, never in the field. Their existence and evolution
are intimately tied to the formation and evolution of the clusters themselves.
The detection of these envelopes, however, is somewhat problematic.

For a time, every elliptical galaxy was thought to have a stellar 
distribution whose projection on the plane of the sky was described by de 
Vaucoulers (1948) $R^{1/4}$ law. This is reflected by the status of "law" 
that is ascribed to what is a highly useful, but nonetheless empirical 
"model". However, Lugger (1984) and Schombert 
(1986) have shown that all luminous, elliptical galaxies, including brightest 
cluster galaxies, have excess flux at large radii relative to their 
best-fitting $R^{1/4}$ models. Moreover, today, 
it is known that only elliptical 
galaxies with $M_B \sim -20.5$ mag have $R^{1/4}$ profiles (e.g. Kormendy 
\& Djorgovski 1989). Brighter and fainter elliptical galaxies are better 
described by S\'ersic's (1963) $R^{1/n}$ 
model (see Graham \& Driver 2005 for a review) 
with the index $n$ taking on values that are greater and 
smaller than 4, respectively (e.g. Caon, Capaccioli \& D'Onofrio 1993; 
Young \& Currie 1994; Graham et al.\ 1996;
Graham \& Guzm\'an 2003, and references therein).  This then leads to the 
question as to whether the excess flux observed in cD galaxies 
(e.g.\ Feldmeier et al.\ 2002; Liu et al.\ 2005) is due to a 
distinct and separate halo of material, or is instead a manifestation from 
the application of, or at least comparison with, an inappropriate fitting 
function, namely the $R^{1/4}$ model.

\begin{table*}
\caption{Column 1: galaxy name. Column 2: host cluster name. Column 3:
  richness class from the catalogue of Abell (1958). Column 4: Type from Bautz
  \& Morgan (1970). Column 5: galaxy redshift taken from the NASA/IPAC
  Extragalactic Database (NED). Column 6: physical scale, i.e.\ the distance
  in kpc that is equivalent to an angular distance of 1$^{\prime\prime}$,
  calculated using a Hubble constant, $H_0=70$ km s$^{-1}$ Mpc$^{-1}$. Columns
  7 and 8: right ascension and declination of the galaxy. Columns 9 and 10:
  right ascension and declination of the nearby blank field. Column 11: worst
  seeing of each galaxy image. Column 12: radial extent of the data.}
\begin{tabular}{llcccccccccc}
\hline
Galaxy         & Cluster    & R.C.     & BM      & z   & Scale       & RA            & Dec            & RA           & Dec              & Seeing     & Extent \\
               &            &          & Type    & & (kpc/$^{\prime\prime}$)
               & \multicolumn{2}{c}{(J2000)}   & \multicolumn{2}{c}{(J2000)}
               & (arcsec) & of data \\
               &            &          &         &    &                              & \multicolumn{2}{c}{Galaxy}    & \multicolumn{2}{c}{Blank field} &    & (kpc) \\
    1          &     2      &    3     &    4    & 5  &        6      &      7    &       8        &      9       &      10         &     11      & 12  \\
\hline
GIN 478        & Abell 2148 & 0        & --           & 0.090 & 1.73  & 16:01:13.9   & +25:27:13      & 16:13:15.4   & +25:28:43        & $\sim$1.2 & 223 \\
NGC 3551       & Abell 1177 & 0        & I            & 0.032 & 0.62  & 11:09:44.4   & +21:45:32      & 10:57:40.2   & +21:47:01        & $\sim$1.3 & 102 \\
NGC 4874       & Abell 1656 & 2        & II           & 0.024 & 0.47  & 12:59:35.7   & +27:57:34      & 12:56:30.9   & +27:53:45        & $\sim$1.5 & 61 \\
NGC 6173       & Abell 2197 & 0        & II-III       & 0.029 & 0.57  & 16:29:44.9   & +40:48:42      & 16:41:52.3   & +40:47:53        & $\sim$2.2 & 94 \\
UGC 9799       & Abell 2052 & 0        & I-II         & 0.034 & 0.67  & 15:16:44.5   & +07:01:17      & 15:28:20.5   & +07:00:53        & $\sim$2.4 & 96 \\
\hline
\end{tabular}
\end{table*}

To address this question, and to derive both the size and 
flux ratio of any possible 
outer envelope relative to the inner galaxy component, obviously one 
should not apply the $R^{1/4}$ model to the inner light-profile.
It 
would similarly 
be a mistake to simply fit an $R^{1/4}$ model to any suspected outer 
halo. We have therefore set out to {\em measure} the shape of the projected 
stellar distribution through the simultaneous application of two $R^{1/n}$
models to the light-profiles from deep exposures of galaxies reported to 
be cD galaxies.

In this paper we analyse $R$-band 
images for 5 cD galaxies observed to a depth of
$\mu_{R}=26.5$ mag arcsec$^{-2}$, at the 3 $\sigma$ level,
with the main purpose of determining the shapes
of the surface brightness profiles of their low surface brightness 
stellar envelopes. From this we also determine the 
galaxy-to-envelope size ratios and the envelope-to-total
flux ratios within 300 kpc, and also when 
applying no outer truncation
to the best fitting models\footnote{The radial extent of our data is typically 
$\sim100$ kpc, and so we have to extrapolate to the truncation radius.}. 
These parameters can then be used to constrain
models of cD galaxy, and host cluster, formation.
This paper is arranged as follows. Section 2 describes
our observations and data reduction, including the ellipse fitting and
surface brightness profile fitting method; 
Section 3 presents the results of the
surface brightness profile fitting and discusses the best fit in each case.
In Section 4 we provide a brief review on the formation process of BCGs and
then go on to discuss our findings. Finally, in Section 5 we summarise our 
main results.

%  NOTE TO EDITOR:  This panel of six figures should be presented in a single column.
\begin{figure*}
\includegraphics{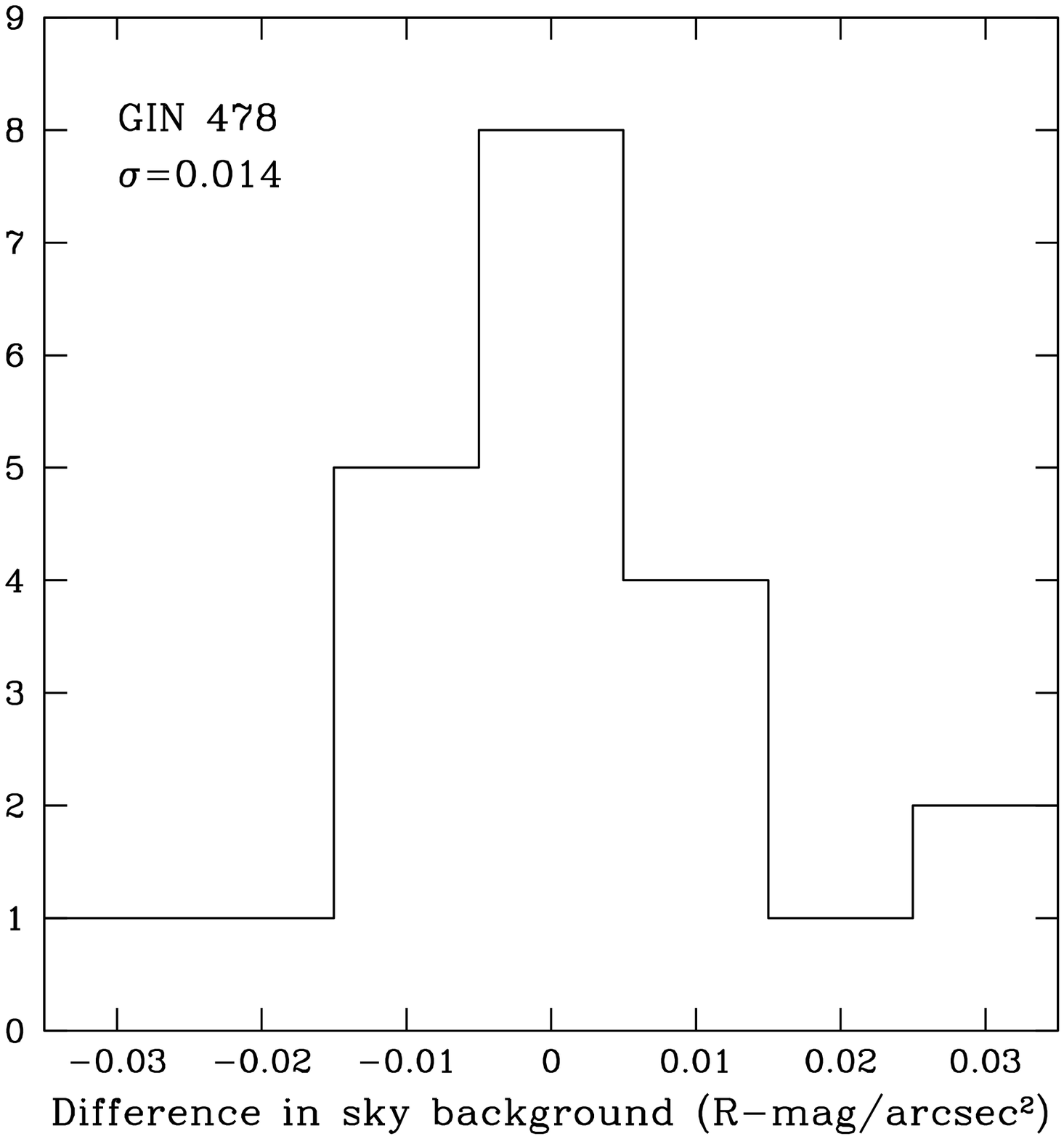}
\includegraphics{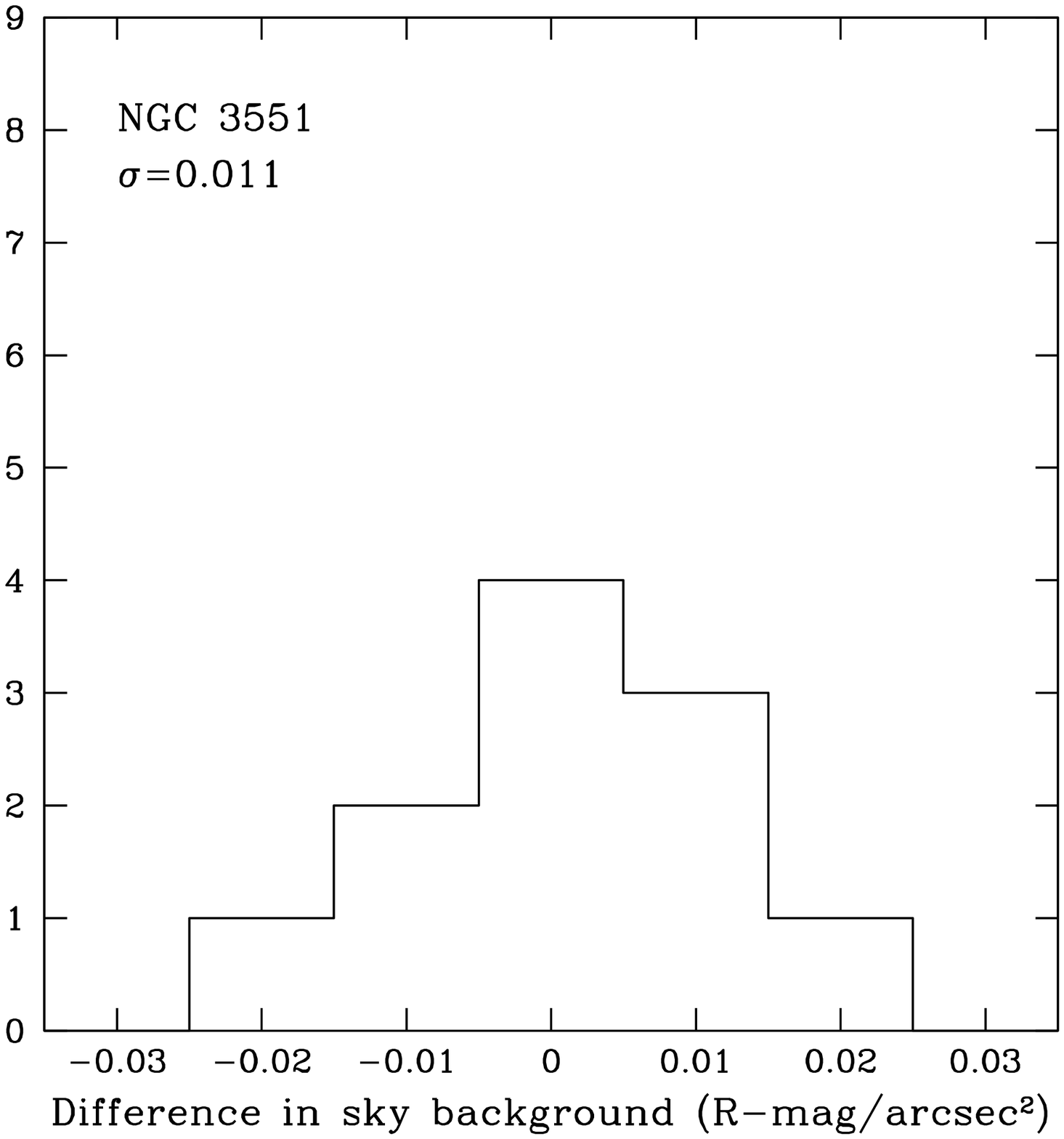}
\includegraphics{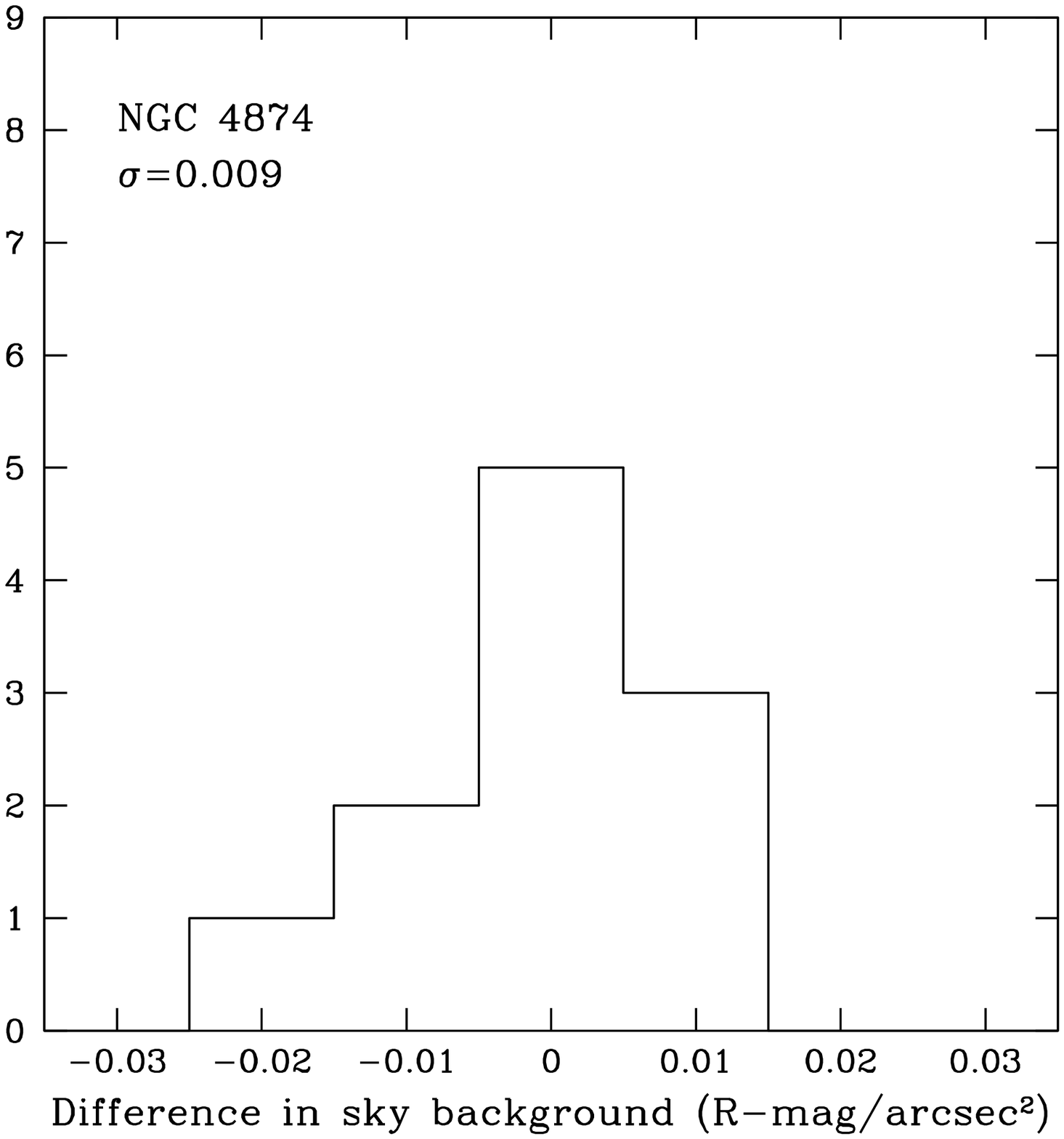}
\includegraphics{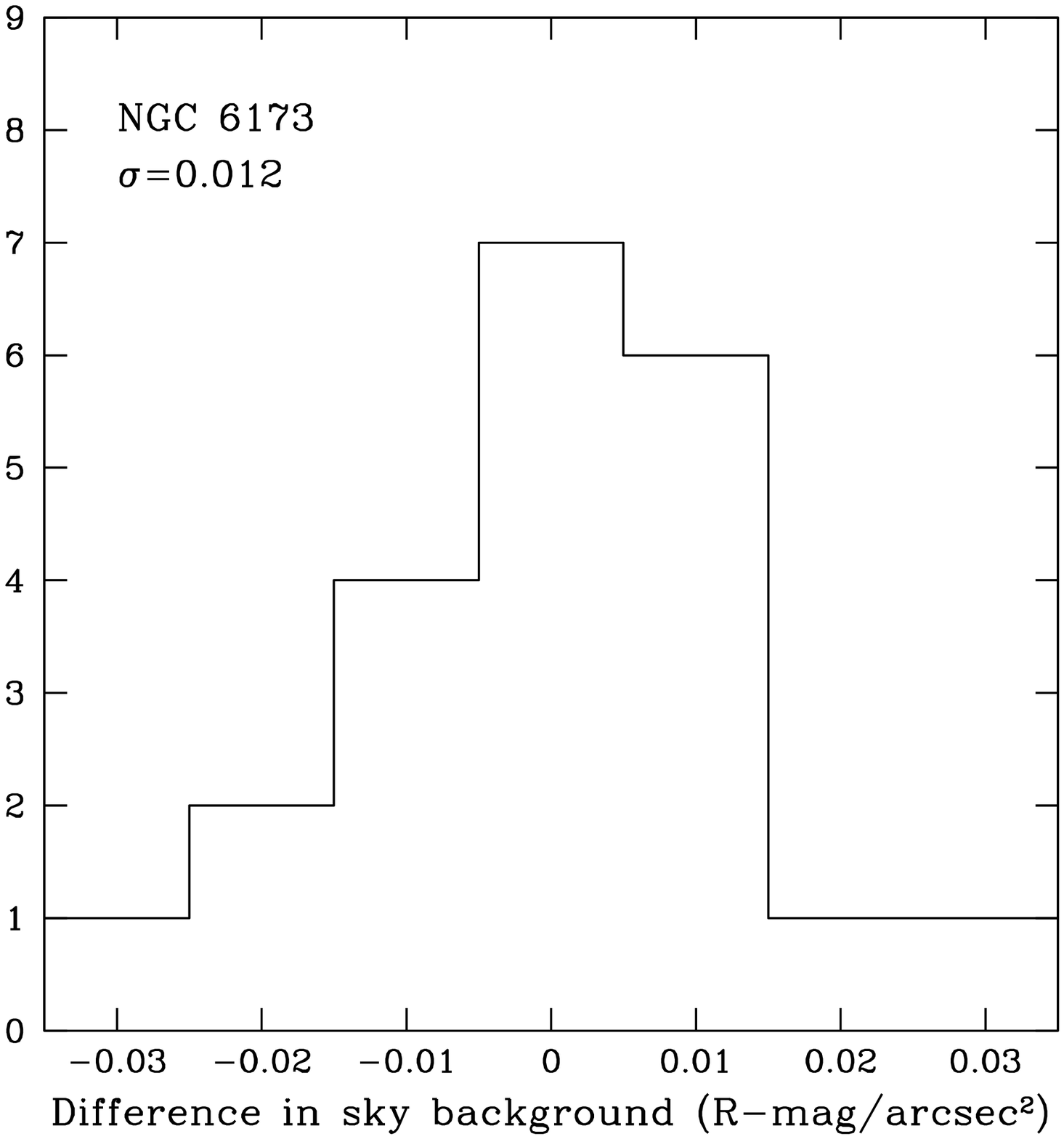}
\includegraphics{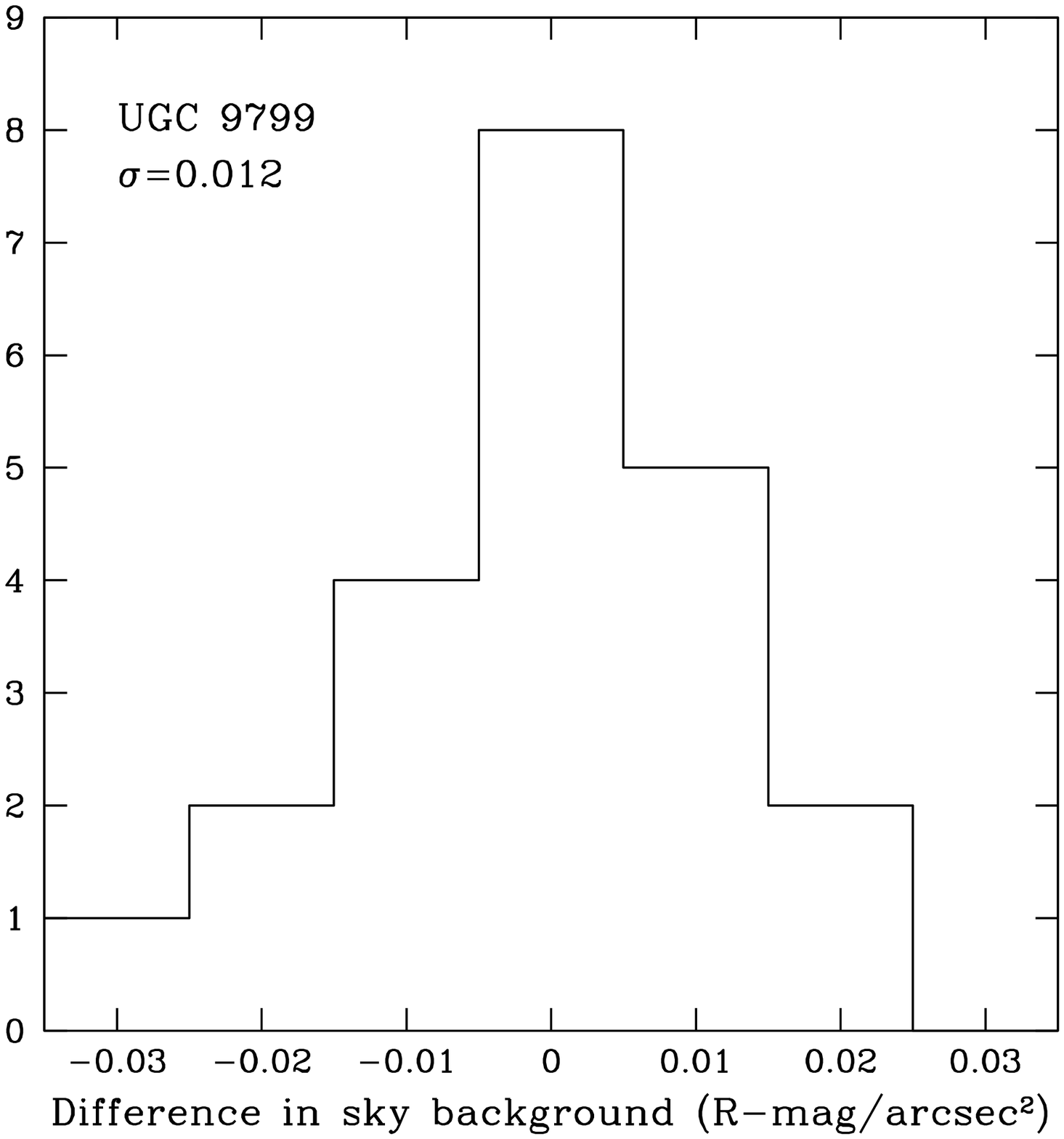}
\includegraphics{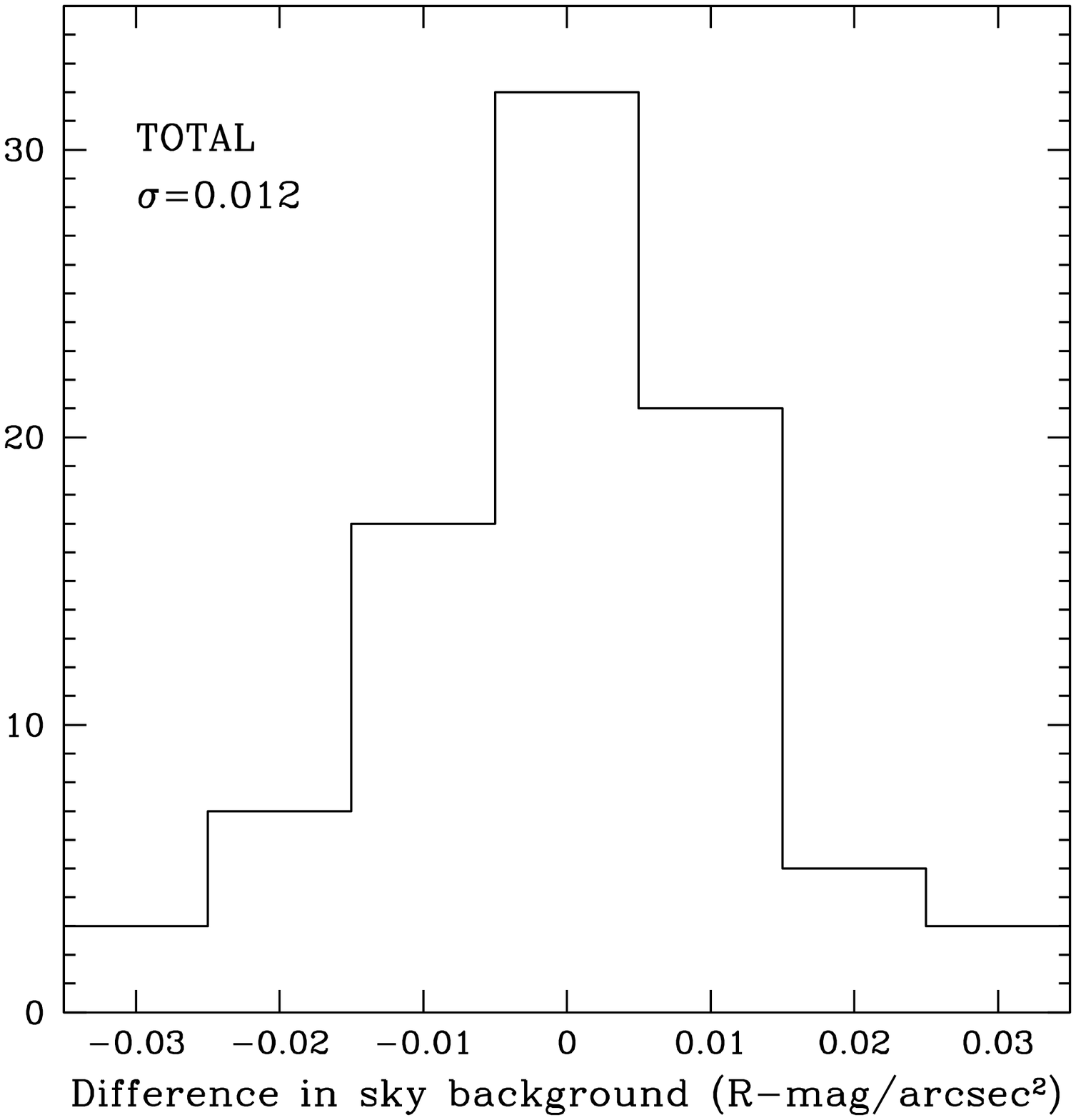}
\vspace*{22cm}
\caption{Differences in the sky background, $\delta_i$, obtained from
  consecutive sky frames near each galaxy, and for the total dataset (bottom
  right panel).  Each panel shows the associated sample standard deviation
  $\sigma = \sqrt{\sum_i^n \delta_i^2 / (n-1)}$, where $n$ is 11, 22 or 88
  depending on the panel.}
\label{sky}
\end{figure*}

\begin{figure*}
\includegraphics{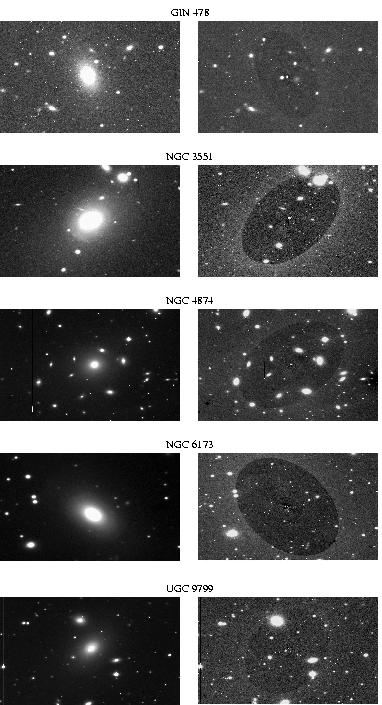}
\vspace*{22cm}
\caption{{\em Left}: Images of the five galaxies in our sample. 
{\em Right}: Residual images derived by subtracting the 
azimuthally-symmetric (galaxy + envelope) model from the original images.}
\label{resimages}
\end{figure*}

\begin{figure*}
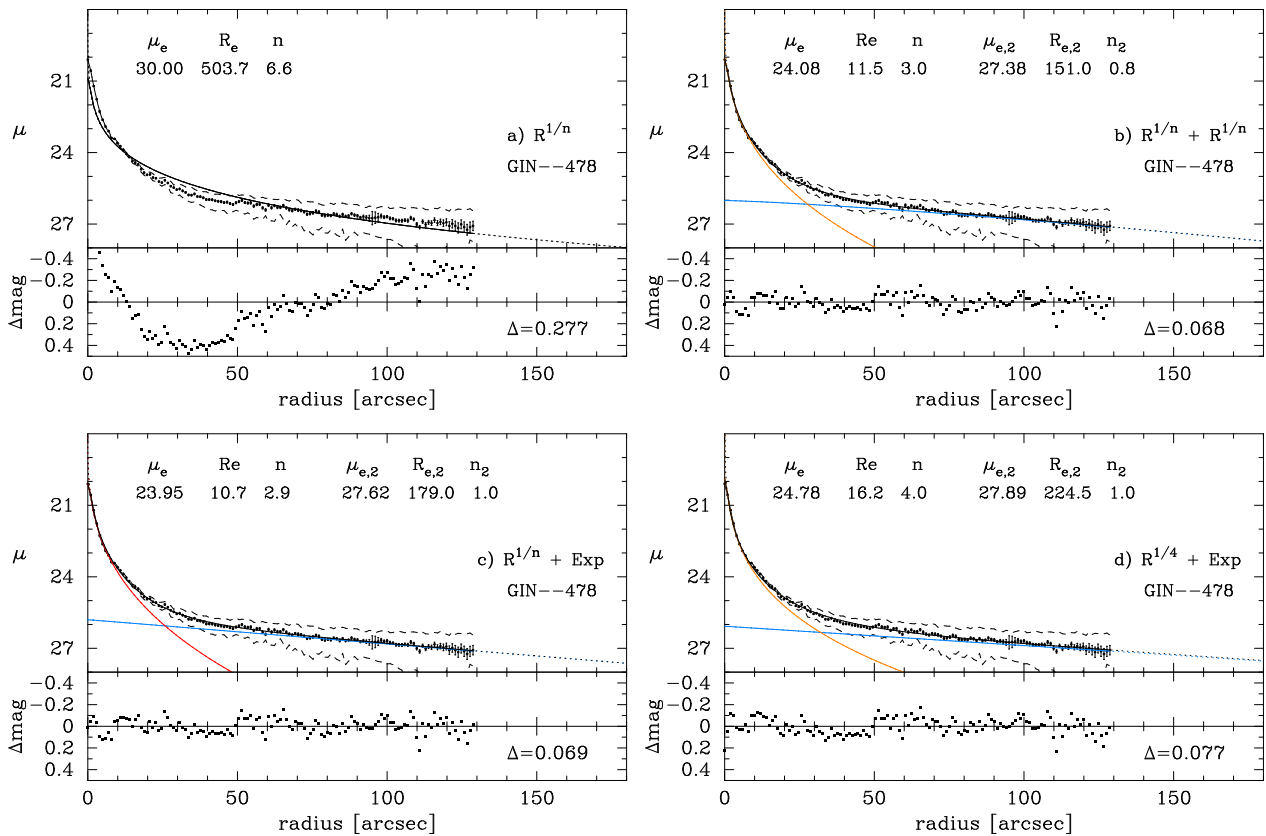

\includegraphics{seigar_fig3a.ps}
\includegraphics{seigar_fig3b.ps}
\includegraphics{seigar_fig3c.ps}
\includegraphics{seigar_fig3d.ps}
\vspace*{11cm}
\caption{Surface brightness profile for GIN 478 with different analytic fits: 
(a) S\'ersic model;
(b) double S\'ersic model; 
% (inner S\'ersic model shown in orange, outer component shown in blue); 
(c) S\'ersic + exponential model; 
(d) $R^{1/4}$ + exponential model. 
Open circles at small radii are 
excluded from the fits. The root mean square (rms) scatter, $\Delta$, is 
shown in the lower portion of each figure. 
The dashed lines indicate the extracted
surface brightness profile with the sky uncertainty added and subtracted.
We note that an arbitrary upper limit of $\mu_{\rm e}=30$ mag arcsec$^{-2}$ in our
software is reached in panel (a), reflecting the inadequacy of a 
single $R^{1/n}$ function to describe the observed stellar distribution.}
\label{gin478}
\end{figure*}

\begin{figure*}
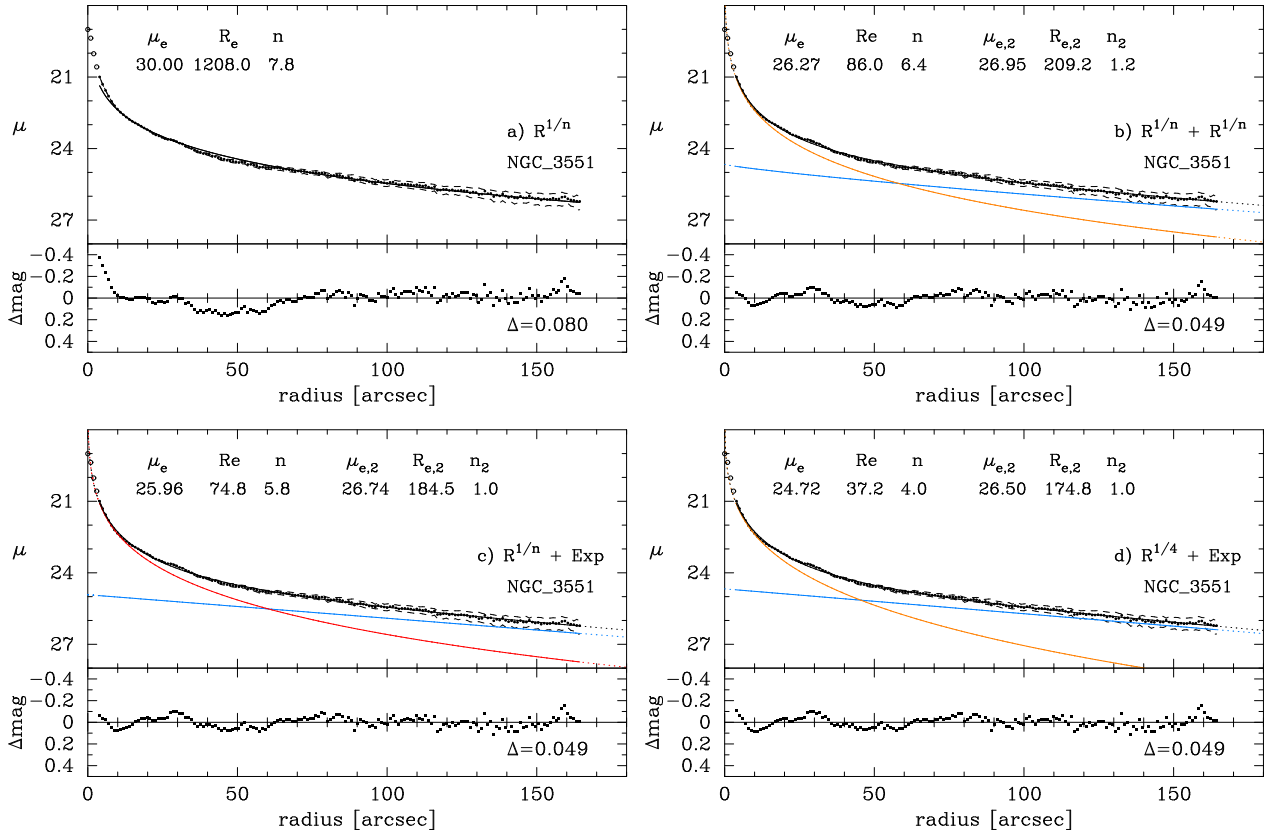

\includegraphics{seigar_fig4a.ps}
\includegraphics{seigar_fig4b.ps}
\includegraphics{seigar_fig4c.ps}
\includegraphics{seigar_fig4d.ps}
\vspace*{11cm}
\caption{Same as Figure~\ref{gin478}, but for NGC 3551.} 
\label{ngc3551}
\end{figure*}

\begin{figure*}
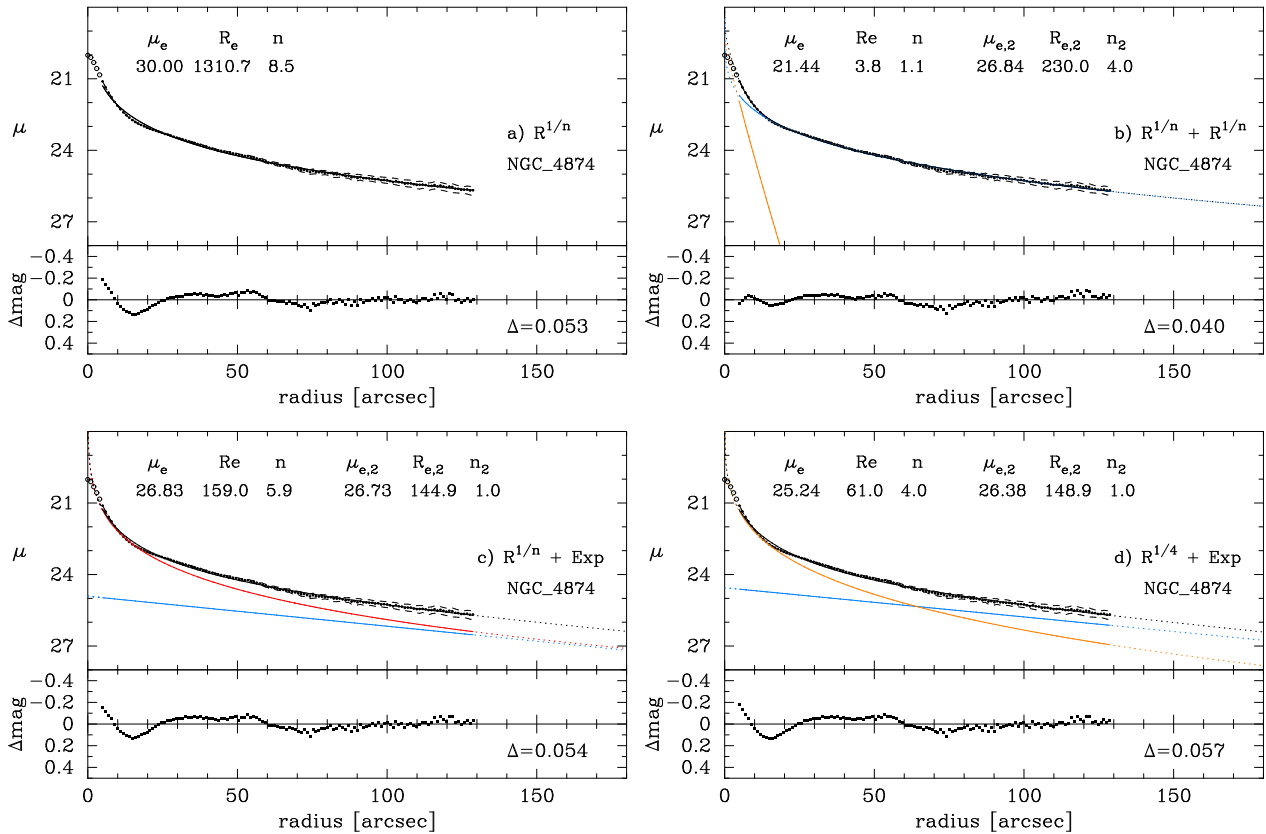

\includegraphics{seigar_fig5a.ps}
\includegraphics{seigar_fig5b.ps}
\includegraphics{seigar_fig5c.ps}
\includegraphics{seigar_fig5d.ps}
\vspace*{11cm}
\caption{Same as Figure~\ref{gin478}, but for NGC 4874.} 
\label{ngc4874}
\end{figure*}

\begin{figure*}
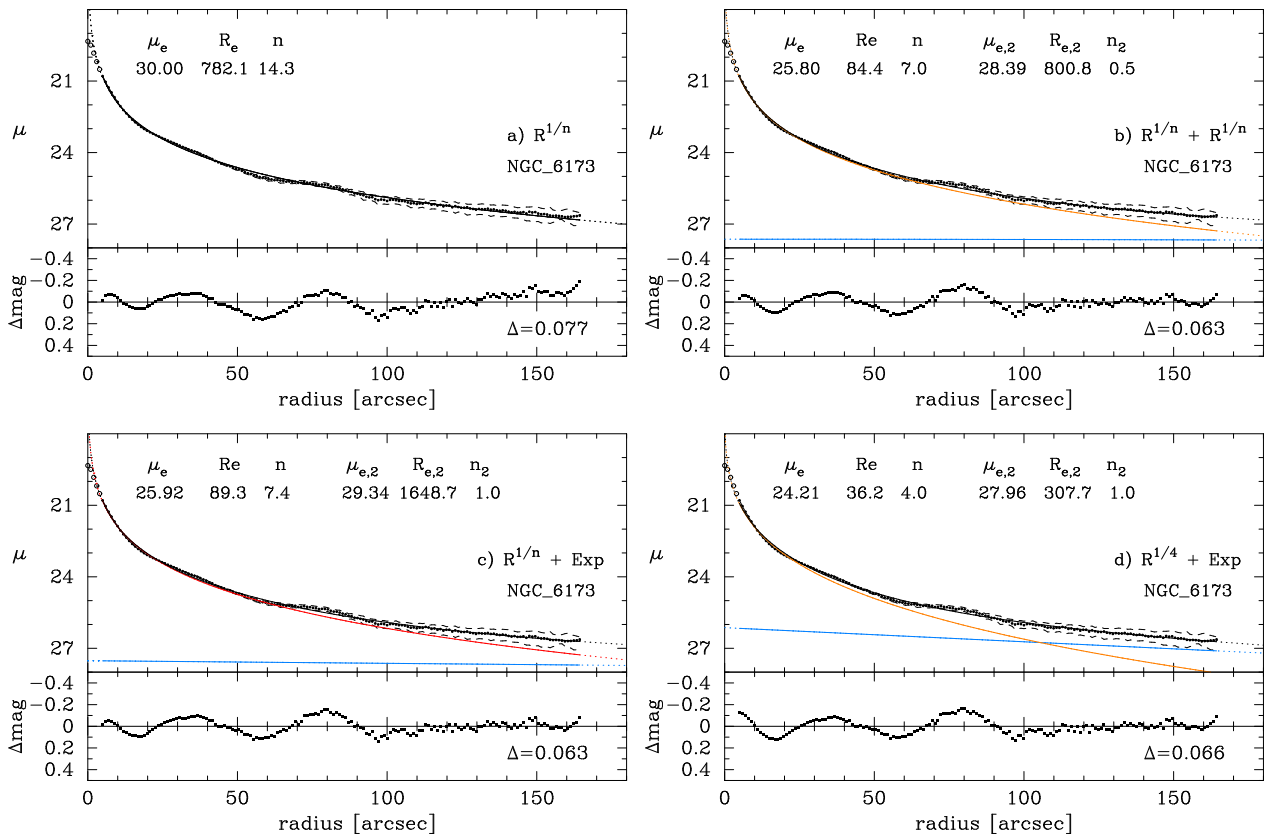

\includegraphics{seigar_fig6a.ps}
\includegraphics{seigar_fig6b.ps}
\includegraphics{seigar_fig6c.ps}
\includegraphics{seigar_fig6d.ps}
\vspace*{11cm}
\caption{Same as Figure~\ref{gin478}, but for NGC 6173. 
The high value of $R_{\rm e, 2}$ in panel (b) is not
a software limit. It is instead an indication that we may be merely
fitting an inadequately subtracted sky-background with the outer $R^{1/n}$
function.}
\label{ngc6173}
\end{figure*}

\begin{figure*}
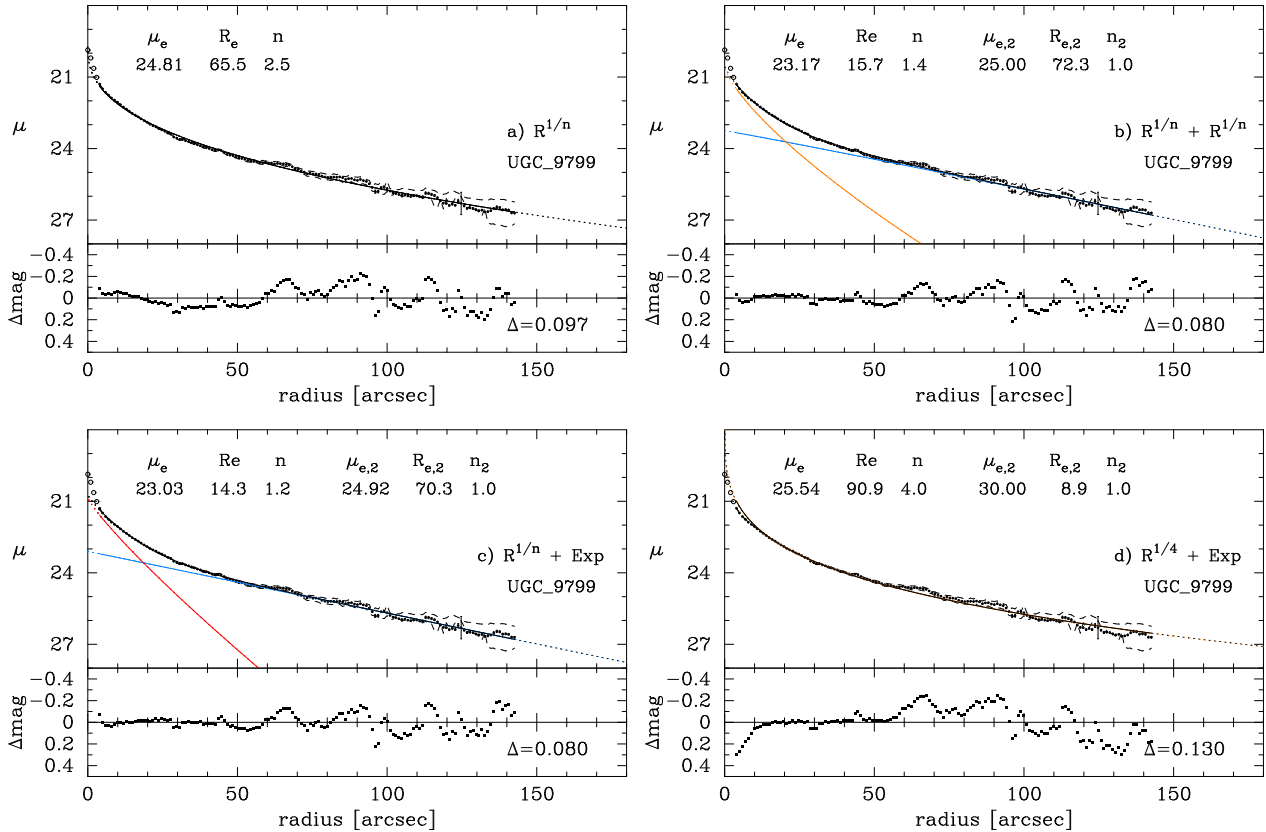

\includegraphics{seigar_fig7a.ps}
\includegraphics{seigar_fig7b.ps}
\includegraphics{seigar_fig7c.ps}
\includegraphics{seigar_fig7d.ps}
\vspace*{11.1cm}
\caption{Same as Figure~\ref{gin478}, but for UGC 9799. In panel (d), the
outer exponential component does not contribute to the fit, and is
thus not seen. The excess central flux seen in panels (b) and (c)
are from the AGN.} 
\label{ugc9799}
\end{figure*}

\section{Observations and Data Reduction}

\subsection{Data}

The light profiles presented in this paper are derived from broad $R$-band 
images of 5 cD galaxies
observed to a depth of $\mu_{R}=26.5$ mag arcsec$^{-2}$ at the 3 $\sigma$ level.
 For elliptical galaxies, Worthey (1994) reports that $B-R=1.65$ 
and Lauer \& Postman (1994) find $B-R=1.51$, and so our surface brightness
limit is equivalent to a $B$-band depth of $\simeq28$ 
mag arcsec$^{-2}$. This is $\sim$3 mag arcsec$^{-2}$ deeper than the study
of BCGs presented by Graham et al.\ (1996).
Our galaxy sample was selected from two samples of BCGs: 
one observed by Lauer \& Postman (1994), the other by Hill \& Oegerle (1993). 
The selection criteria for these galaxies was that they (i) were classified
as cD galaxies, (ii) had a limiting redshift $z<0.1$, and (iii) were visible
from La Palma in late April.  Of the galaxies that
met the selection criteria, we observed five; they are listed in Table 1.

On each night at least four standard stars from the Landolt (1992) list were
observed, at varying airmass, in order to determine the photometric zero-point 
and the airmass extinction correction for the data. These have been applied to 
each image. Corrections for surface brightness dimming, galactic extinction
and K-correction have also been applied. Throughout this paper, magnitudes are
quoted using the Vega system.

For each galaxy, one nearby blank field (see Table~1)
was also observed for the purpose of estimating the sky-background free from
any intracluster light. The observations were taken in such a way that the
object frames and sky frames were interleaved.
The observations were made during dark time 
with the 1.0-m Jacobus Kapteyn Telescope (JKT) on
the island of La Palma. These observations were taken on the nights of 2003 
April 24--30 using the 2048$\times$2048 SITe2 CCD camera, which has 0.331 
arcsec pixels and a field-of-view
11.3$\times$11.3 arcmin$^{2}$. The filter used
was a standard Harris $R$ band filter. 
The observations of both the galaxy fields and the blank fields consisted
of 24 co-added exposures of 900 seconds. The exceptions to this are NGC 3551 
and NGC 4874, where only 12 co-added exposures were made due to bad weather.

Data reduction was performed within $\tt IRAF$. All images were 
bias-subtracted, and then flat fielded using twilight flats. Images
were then combined by degrading the best images, by convolution with a
Gaussian, to match the seeing conditions in the worst image, 
which was typically 
$\sim 1^{\prime\prime} - 2^{\prime\prime}$ (see Table 1). There was no
evidence for any fringing effects in any of the images. The blank sky images
were then used to estimate the sky-background and these values were subtracted 
from the galaxy images. In performing the sky-background estimation, ten
areas of the 
blank sky image were used to calculate 10 medians. The sky
value was then taken as the mean of these 10 medians, and the average deviation
was adopted as the uncertainty in the sky background. 
% The sky-background value and its uncertainty are shown in Table~\ref{xxx}. 

Uncertainties in the estimation of the sky-background are the dominant source
of error in determining the shape of the outer part of the galaxy surface
brightness profiles.  The above estimates of this error have been used in
Section~2.3 to quantify the variation in the optimal parameters of the fitted
models.  Here we adopt an alternative approach to gauge the size of this
error.  Because of the way we interlaced our target and sky frames, the
sky-background in each such pair of observations may differ.  In
Figure~\ref{sky} we show histograms of the difference in the sky background
between consecutive sky frames for each galaxy.  Each offset in these
histograms roughly represents twice the expected background offset between an
individual target and sky frame, for which only half as much time has elapsed
between their acquisition.  As can be seen, the median value for this
difference is consistent with the sky frames having similar values for the
background.  
The sample standard deviation, $\sigma$, is only $\sim$0.012 mag 
arcsec$^{-2}$. A dominant monotonic drift in the background as one 
progresses through the night would produce a non-zero median value in the 
distributions to either positive or negative values. 
Given $N$ ($=$11 or 
22) pairs of sky frames, the expected difference in the final 
sky-background level of each is given by $\sigma/\sqrt{N}$. This amounts 
to 0.0026--0.0036 mag arcsec$^{-2}$ uncertainty, consistent with the sky 
errors calculated using the method above.

\subsection{Extraction of surface brightness profiles} \label{SecEx}

We have employed and compared two methods for the extraction of the surface 
brightness profiles from the galaxy images. 

The first is isophotal
ellipse fitting, performed using the {\bf Ellipse} routine
in {\tt IRAF} which uses
an iterative method described by Jedrzejewski (1987). 
Each isophote was fitted allowing for a variable position angle and ellipticity,
but holding the centre fixed.
Foreground and background sources were masked out within the
{\bf Ellipse} routine. 
Given our interest in the outer light-profile, we chose to sample the 
surface brightness profile using a standard, albeit somewhat arbitrary,
linear spacing. A logarithmic spacing would have generated a lot of data
points, and hence more weight, at small radii. % (e.g.\ Graham 2002). 
The {\bf Ellipse} routine works in such
a way, that ellipses are set up centered on the central cD galaxy. 
This defines several elliptical annuli. The surface
brightness at each annulus is calculated as the average number of counts
within that annulus. The uncertainty on the surface brightness
is calculated as the rms between pixels divided by $\sqrt{N-1}$, where $N$
is the number of pixels. The statistical error is therefore dependent upon
the step size between annuli or ellipses. However, at large radii, 
the actual error on
the computed surface brightness, is dominated by uncertainty in the sky
background, and we discuss this later.

The second method that we used determines the surface brightness profiles
with the help of another isophotal fitting routine also written in {\tt IRAF}
(Jerjen, Kalnajs \& Binggeli 2000; Barazza, Binggeli \& Jerjen 2002). 
After foreground stars and
neighbouring galaxies were removed from the image, a symmetrical 2-D model was
reconstructed from the observed light distribution allowing isophotal
ellipticity and position angle to vary with radius, 
but keeping the luminosity-weighted
centre fixed. This fitting process was repeated iteratively until the residuals
were minimised. The 1-D surface brightness profiles were
then calculated from the 2-D model by adopting mean values for ellipticity
and position angles. 
% (not done)  
% In this case, surface brightness profiles with the 1 $\sigma$ sky-background 
% uncertainty added and subtracted from the image have also been derived. 

Figure \ref{resimages} shows the reduced images (left panel) and the
residual images (right panel) of the 5 cD galaxies in our sample. The residual 
images were produced by subtracting
an azimuthally symmetrical model from the original
data.

Both sets of surface brightness profiles, and the subsequent modeling,
generated consistent results. From here on we refer to 
only the surface brightness profiles
extracted using the {\tt IRAF} task, {\bf Ellipse}, presented
in Figures \ref{gin478} to \ref{ugc9799}. The dashed lines in Figures 
\ref{gin478} to \ref{ugc9799} represent the
extracted surface brightness profile after adding (upper dashed line)
and subtracting (lower dashed line) the uncertainty in the sky background.
Note that this uncertainty was added to or subtracted
from the image, and then the new surface brightness profile was derived.
Because this uncertainty is applied to the image, the dashed lines sometimes
do not necessarily fall either side of the data points. This is because it is
possible for the {\bf Ellipse} routine to fit the image with slightly
different ellipticities and position angles, and so a small adjustment
in the surface brightness results.

\subsection{Modelling the surface brightness profiles}

Our modelling of the surface brightness profiles employs 
the S\'ersic (1963, 1968) 
model for both the inner part of the galaxy and the outer stellar envelope. 

The S\'ersic  $R^{1/n}$ radial intensity profile can be written as
\begin{equation}
I(R)=I_{e}\exp{\left\{-b_{n}\left[\left(\frac{R}{R_{\rm e}}\right)^{1/n}-1\right]\right\}},
\end{equation}
where $I_{e}$ is the intensity at the effective radius, $R_{\rm e}$, which
encloses 50 per cent of the light. The factor $b_{n}$ is a function of the 
shape parameter, $n$, such that $\Gamma(2n)=2\gamma(2n,b_{n})$, where $\Gamma$
is the gamma function and $\gamma$ is the incomplete gamma function
(see Graham \& Driver 2005). 
In the case where $n=1$, the S\'ersic model is equivalent to an 
exponential, and when $n=4$ it is equivalent to the $R^{1/4}$ model.

Initially, for all the
galaxies, we attempt to fit the entire surface brightness profile with a 
single S\'ersic component.
We also test the applicability of fitting an inner $R^{1/n}$ model and
an exponential model to the outer envelope (i.e., in a similar way to
the case for disk galaxies, e.g., Andredakis et al.\ 1995; 
Seigar \& James 1998; Graham 2001).
We then go on to use a multitude of fits (not all shown), which keep the
inner and outer S\'ersic indices fixed at integer values between 
1 and 4. Finally, we allow both of the S\'ersic indices to
vary. 

Corrections for the effects of seeing have been made using the prescription
given in Pritchet \& Kline (1981). Due to the Gaussian nature of the JKT 
point spread function (psf)
it is not necessary to consider more complicated seeing corrections. 
For any intrinsically radially symmetric
intensity distribution, $I(R)$, the observed seeing-convolved profile, 
$I_{c}(R)$, is
\begin{equation}
I_{c}(R)=\sigma^{-2}e^{-R^{2}/2\sigma^{2}}\int_{0}^{\infty}I(x)I_{0}(xR/\sigma^{2})e^{-x^{2}/2\sigma^{2}}xdx,
\end{equation}
where $\sigma$ is the dispersion of the Gaussian psf,
which is equal to the full-width half maximum (FWHM) divided by a factor
of 2.3548. The $I_{0}$ term is 
the zeroth-order modified Bessel function of the 
first kind (e.g. Press et al.\ 1986). This approach to correcting light profile 
shapes for seeing was adopted by Andredakis, Peletier \& Balcells (1995) and 
later by de Jong (1996). 

Due to the potential presence of partially depleted cores in luminous 
elliptical galaxies (e.g., Lauer et al.\ 1995; Graham et al.\ 2003; Trujillo et 
al.\ 2004; Ferrarese et al.\ 2006 and references therein), 
or instead the presence of multiple nuclei from semi-digested 
mergers, the innermost seeing-effected data points ($\sim3^{\prime\prime}, 
\sim2-5$ kpc) have been excluded from the fits. In comparison, the inner 
10-20 kpc were excluded from the BCG analysis in Zibetti et al.\ (2005), where 
the FWHM was $\sim$5 kpc. It is because of such features that one should 
not model integrated aperture magnitude profiles, 
in which every data point is 
effected/biased. Instead, one should fit the surface brightness profiles 
directly.
Because of the AGN in UGC 9799, the inner 4 data points were excluded.
Obviously, given that we have excluded the most seeing affected data, the
use of equation 2 to convolve our $R^{1/n}$ models before fitting them to the
observed light profiles is not so crucial. Deactivating the seeing correction
has no significant affect.

The best-fitting models were acquired using the
subroutine UNCMND from  Kahaner, Moler \& Nash (1989).
At each iteration, the nonlinear S\'ersic functions are approximated
by a quadratic function derived from a Taylor series.
The quadratic function is minimised to obtain a search direction,
and an approximate minimum of the nonlinear function along
the search direction is found using a line search.  The
algorithm computes an approximation to the second derivative
matrix of the nonlinear function using quasi-Newton techniques.

Common practice when fitting a model to some data set is to
employ the use of the $\chi^2$ statistic, such that the reduced-$\chi^2$
is given by,
\begin{equation}
\chi^2 = \frac{\sum_{i=1}^m \delta_i^2/\sigma_i}{m-k},
\label{Eq_chi}
\end{equation}
where $m$ is the number of data points, $\delta_i$ is the $i$th
residual (about the best-fitting model), 
$\sigma_i$ is the uncertainty on the $i$th data point
and $k$ is the number of parameters in the fitted model.

Such an approach is highly desirable \textit{if} one knows what the
functional form of the underlying model is.  However, when one does
not know the form of the underlying model, but instead has to assume
some empirical function such as de Vaucouleurs model or S\'ersic's
model, the use of the reduced-$\chi^2$ statistic can produce rather biased
results. For example, at the centre of a galaxy, the signal-to-noise ratio is
high, and thus the uncertainties ($\sigma_i$) that are assigned to the
central data are small.  These points thus have considerable weight in
determining the best fit, obtained by minimising the 
reduced-$\chi^2$ value.
In the past, the reduced-$\chi^2$ statistic has been used to fit S\'ersic
bulges plus exponential disks to spiral galaxy light profiles.
However, due to the presence of additional (un-modelled) nuclear
components, the bulge model and the simultaneously-fit disk model have
been heavily biased (see Balcells et al.\ 2003).  That is, because the
assumed model (bulge + disk) 
did not match the true underlying distribution 
(bulge + a disk + an additional nuclear component), the
small uncertainties on the data at small radii heavily biased the fits
to produce erroneous results (see e.g., Schombert \& Bothun 1987).

We do not know the structural make-up of
our 5 cD galaxies, they may contain a third component, such as a bar
or a lens or indeed multiple nuclear components that we do not model.
We therefore wish to avoid use of the $\chi^2$ statistic.

Furthermore, another main concern is that we wish to quantify
the stellar distribution of
the suspected outer envelope.  In using 
equation \ref{Eq_chi}, the (signal-to-noise)-weighted values of
$\sigma_i$ will act to erase the value or worth of the data at large
radii.  Moreover, in trying to gauge the influence of sky-background
errors, likely to be a major source of uncertainty on the shape of the
outer stellar distribution, the use of equation \ref{Eq_chi} would
dilute the effect of adding and subtracting the uncertainty in the
sky-background, and give one the false belief that their fitted models
have less variance than they really ought. This is because correlated
errors are not taken into account when computing the $\chi^2$ value. 

A common approach, which circumvents the above two problems, 
and which we have adopted, is
to use the root mean square (rms) scatter
\begin{equation}
\Delta = \sqrt{\frac{\sum_{i=1}^m {\delta_i}^2}{m}}.
\label{Eq_Delta}
\end{equation}

The results of our profile fitting are shown in Figures~\ref{gin478}--\ref{ugc9799}. The 
S\'ersic indices, effective radii and effective surface brightnesses are 
listed in Table 2. 
Uncertainties on the best-fitting parameters (Table 2) are obtained by
repeating the fit to the surface brightness profiles after adding and
subtracting the 1 $\sigma$ uncertainty in the sky-background level. It
should be noted that the effective radii are model parameters that
provide the optimal fit to the data, over the observed data range.
They additionally reflect real, physical half-light radii only if the
models can be extrapolated to infinity.

While the introduction of additional free 
parameters in a fitted model will reduce the
value of $\Delta$, we show here that this is not the explanation for the
improvement in the fit we obtain when changing from an $R^{1/4}$ model 
to an $R^{1/n}$ model.

Our profiles have at least $m=125$ measured points. $R^{1/4}$ +
exponential fits have four free parameters, i.e. $k=4$, and
S\'ersic + S\'ersic fits have $k=6$. For random residuals, the expected value of $\Delta$
scales as $\sqrt{(m-k)/m}$. As a result, an improvement of only
0.8\% would be expected for increasing the number of free 
parameters from 4 to 6. To have an improvement of 5\% in $\Delta$, a total
of 16 free parameters are needed, and for a 10\% improvement,
27 free parameters are needed. For three of our light profiles 
an improvement
of more than 10\% is shown when adopting a S\'ersic + S\'ersic fit
over an $R^{1/4}$ + exponential fit.  That is, the residuals
about the $R^{1/4}$ + exponential fit are not random, instead, there is 
structure indicating the inadequacy of this model and justifying 
the double S\'ersic model.  NGC 3551 
shows little, if any, improvement in its $\Delta$, revealing that
an $R^{1/n_1}+R^{1/n_2}$ with $n_1=4$ and $n_2=1$ is appropriate
for this galaxy. A summary of the values of $\Delta$ found for
each type of fit applied to each of our galaxies is shown in Table
\ref{delta}. We do not consider NGC 6173
here since only one component is necessary to model
this galaxy (see Section 3).

\begin{table*}
\caption{
Summary of the results of the double S\'ersic model fitting, except NGC 6173
for which we present the results from the single S\'ersic component
fit. Column 1: galaxy name. Column 2: total absolute magnitude within a 300
kpc radius. Column 3 and 4: surface brightness $\mu_{\rm e}$ at the effective
radius $R_{\rm e}$, respectively, from the inner S\'ersic fit. Column 5: inner
S\'ersic index $n_1$. Column 6: absolute magnitude within a 300 kpc radius for
the inner component. Column 7 and 8: surface brightness at the effective
radius, and this radius, for the outer S\'ersic component. Column 9: outer
S\'ersic index $n_2$. Column 10: absolute magnitude within a 300 kpc radius 
for the outer component. Given that the dominant source of error in the 
observed surface brightness profiles arises from uncertainties in the 
sky-background, all parameter errors are obtained by re-fitting the models to
the images with the sky uncertainty added or subtracted.}
\label{Tab2}
\begin{tabular}{llllllllll}
\hline
Galaxy         & $M_{tot}$       & \multicolumn{4}{c}{Inner}                                                 & \multicolumn{4}{c}{Outer}                                                 \\
               & $<300$ kpc      & $\mu_{\rm e, 1}$         & $R_{\rm e, 1}$    & $n_1$       & $M_{inner}$          & $\mu_{\rm e, 2}$         & $R_{\rm e, 2}$     & $n_2$       & $M_{outer}$         \\
               &                 & (mag arcsec$^{-2}$)   & (arcsec)      &
               & $<300$ kpc           & (mag arcsec$^{-2}$)   & (arcsec)       &             & $<300$ kpc          \\
    1          &     2      &    3     &    4    & 5  &        6      &      7    &       8        &      9       &      10   \\
\hline
GIN 478        & -23.03$\pm$0.45 & 24.08$\pm$0.13       & 11.5$\pm$0.9  & 3.0$\pm$0.1 & -21.63$\pm$0.03      & 27.38$\pm$0.05       & 151.0$\pm$49.1 & 0.8$\pm$0.1 & -22.69$\pm$0.43      \\
NGC 3551       & -22.21$\pm$0.24 & 26.27$\pm$0.49       & 86.0$\pm$18.7 & 6.4$\pm$0.7 & -21.50$\pm$0.03      & 26.95$\pm$0.23       & 209.2$\pm$49.8 & 1.2$\pm$0.2 & -21.40$\pm$0.22      \\
NGC 4874       & -21.78$\pm$0.09 & 21.44$\pm$0.23       & 3.8$\pm$0.3   & 1.1$\pm$0.2 & -18.50$\pm$0.02      & 26.84$\pm$0.13       & 230.0$\pm$26.1 & 4.0$\pm$0.1 & -21.73$\pm$0.08      \\
NGC 6173       & -22.95$\pm$3.57 & 30.00$\pm$5.47       & 788$\pm$218& 14.0$\pm$7.5& --                 & --                 & --             & --            & --                  \\
UGC 9799       & -22.26$\pm$0.45 & 23.17$\pm$0.17       & 15.7$\pm$3.1  & 1.4$\pm$0.2 & -20.73$\pm$0.19      & 25.00$\pm$0.19       & 72.3$\pm$13.4  & 1.0$\pm$0.3 & -21.95$\pm$0.29      \\
\hline
\end{tabular}
\end{table*}

\begin{table*}
\caption{Summary of the values of $\Delta$ found for each type of fit applied
  to 4 of our 5 galaxies.  See Section~\ref{SecEx}.
}
\label{delta}
\begin{tabular}{lcccccc}
\hline
Galaxy         & \multicolumn{6}{c}{$\Delta$ (mag arcsec$^{-2}$)}  \\
               & $R^{1/n}$  & $R^{1/4}+R^{1/4}$  & $R^{1/3}+R^{1/3}$  & $R^{1/4}+$Exp  & $R^{1/n}+$Exp  & $R^{1/n}+R^{1/n}$  \\
\hline
GIN 478        & 0.277      & 0.118              & 0.076              & 0.077          & 0.069          & 0.068              \\
NGC 3551       & 0.080      & 0.050              & 0.050              & 0.049          & 0.049          & 0.049              \\
NGC 4874       & 0.053      & 0.042              & 0.046              & 0.057          & 0.054          & 0.040              \\
UGC 9799       & 0.097      & 0.130              & 0.105              & 0.130          & 0.080          & 0.080              \\
\hline
\end{tabular}
\end{table*}

Unfortunately neither the ellipticity profile nor
the position angle profile yielded any clues to the transition from
inner to outer component.
Similarly, Zibetti et al.\ (2005, their Fig.6) show that 
no change in the 
ellipticty profile is observed at the inflection of their surface 
brightness profile.  In fact, there is no change in the behaviour of their 
ellipticity profile until $\sim$160 kpc --- a radius 8 times greater than 
their inner component's effective radius.

\section{Results: Galaxy surface brightness profiles} \label{results}

%  NOTE TO EDITOR:  This panel of four figures should be presented in a single column.
\begin{figure*}
\includegraphics{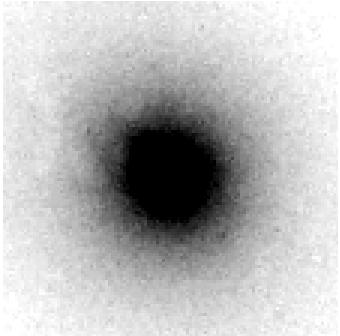}
\includegraphics{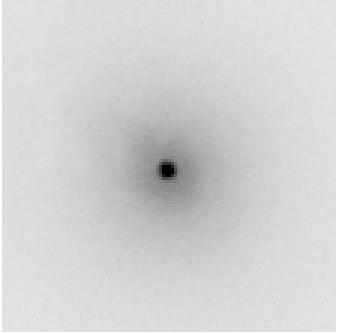}
\includegraphics{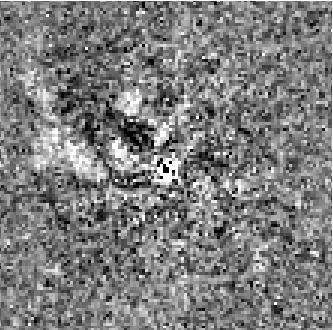}
\includegraphics{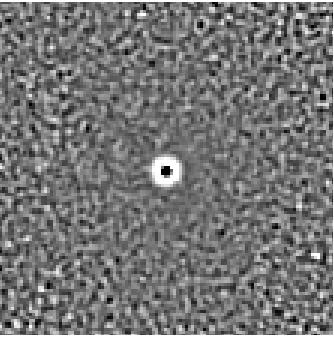}
\vspace*{15.5cm}
\caption{{\tt HST/WFPC2 F814W} images of the inner 
$5^{\prime\prime}$ of UGC 9799. In all images north is up
and east is left. {\em Top left}: A high contrast image
showing signs of fine structure to the north-east of the nucleus
of UGC 9799. {\em Top right}: A low contrast image showing the
AGN. {\em Bottom left}: An unsharp-masked image, highlighting 
the fine structure to the north-east of the nucleus. {\em Bottom
right}: A structure map (see Pogge \& Martini 2002) highlighting 
the AGN.}
\label{struct}
\end{figure*}

\noindent
{\em GIN 478}:
\\

\noindent
The obvious structure in the residual profile of Figure \ref{gin478}a 
reveals that
this galaxy is not well described with a single-component $R^{1/n}$ model.
Figure \ref{gin478}b, in which two $R^{1/n}$ models have been fitted, 
shows that the outer part of this galaxy's surface brightness 
profile is almost exponential in nature. Fitting an outer exponential 
(Figure \ref{gin478}c) 
shows that the profile shape of the best-fitting inner model
does not change significantly, with a S\'ersic index $n \sim 3$. 
Both of these fits have residuals of $\Delta \simeq 0.07$ mag arcsec$^{-2}$. 
Fitting 
the inner part of the profile with an $R^{1/4}$ model and
the outer part of profile with an exponential (Figure \ref{gin478}d) 
increases the residuals by more than 10 per cent with $\Delta=0.077$ mag
arcsec$^{-2}$. An $R^{1/3}$ + exponential model is the optimal fit when 
using integer values for the S\'ersic indices.
\\
\\

\noindent
{\em NGC 3551}:
\\

\noindent
This galaxy is also not well modelled with a single $R^{1/n}$ function 
(Figure \ref{ngc3551}a),
evidenced by the mismatch at small radii and the relatively large rms scatter
$\Delta$ (c.f. Figure \ref{ngc3551}b, c and d).
Figure \ref{ngc3551}b 
shows that the outer part of this galaxy's surface brightness 
profile is almost exponential in nature. Modelling the outer profile with an
exponential shows that the inner $R^{1/n}$ model's profile shape does not
change significantly, with a S\'ersic parameter $n \sim 6$ (Figure 
\ref{ngc3551}c). 
Both of these fits have residuals of $\Delta=0.049$ mag arcsec$^{-2}$.
Figure \ref{ngc3551}d shows an $R^{1/4}$ fit to the inner component and 
an exponential fit to the outer component, which also
has a residual of $\Delta=0.049$ mag arcsec$^{-2}$
A S\'ersic model
with index $n \sim 4-6$ therefore appears to provide a good fit to the central galaxy in this
case, with an outer exponential law again describing the envelope.
The change from an inner S\'ersic parameter of $n=4$ to $n=6.4$ 
results in an increase in $R_{\rm e}$ from 37 to 86 arcsec, i.e.\ more than
a factor of two.  This is the only galaxy for which this degeneracy is
seen, and this is not typical of our galaxies in general.
\\
\\

\noindent
{\em NGC 4874}:
\\

\noindent
The residual profile in Figure \ref{ngc4874}a resembles that seen in 
Figures \ref{gin478}a and \ref{ngc3551}a,
revealing additional structure that a single-component S\'ersic model cannot
describe. Curiously,
Figure \ref{ngc4874}b 
shows that fitting the system with a double S\'ersic
model provides an 
outer S\'ersic index of $n \sim 4$, which is not seen in any of the other 
galaxies. 
Moreover, the inner galaxy seems well characterized by a profile that is
close to exponential.
If we force the outer model to have an exponential profile (Figures 
\ref{ngc4874}c and d), 
the residuals increase significantly
from $\Delta=0.040$ mag arcsec$^{-2}$ to $\Delta=$0.054--0.057 mag arcsec$^{-2}$. 
The inner component of NGC 4874 has a notably small effective radius of
$\sim$4 kpc. This is perhaps not unusual as Gonzalez et al.\
(2005) also have a number of BCGs with $R_{\rm e, inner}<5$ kpc.
\\
\\

\noindent
{\em NGC 6173}:
\\

\noindent
The residual profile in
Figure \ref{ngc6173}a demonstrates that a single-component fit may be
adequate
for this galaxy, although a large S\'ersic index ($n\sim14$)
is required, and the values of $\mu_{\rm e}$ and $R_{\rm e}$ are unreasonably
large. Indeed, the value of $\mu_{\rm e}=30$ mag arcsec$^{-2}$ is an
artificial upper limit used in the fitting code.
Including an outer component decreases the
residual from $\Delta=0.077$ to $\Delta=0.063$ (Figure \ref{ngc6173}b and
\ref{ngc6173}c) and requires an inner S\'ersic index of $n\sim 7$.
However, given the near constant surface brightness of this outer
component over the radial range for which we have data, at first 
glance it looks as if we failed to adequately subtract the sky background.
The {\em additional} subtraction of the 1 $\sigma$ uncertainty
in the sky background results in the disappearance of this near
constant surface brightness component seen in Figures \ref{ngc6173}b, c and d, 
and the optimal fit is actually a 1-component fit, with 
$n=8.7$, $R_{\rm e}=141.2$ arcsec and $\mu_{\rm e}=25.1$ mag arcsec$^{-2}$. We
thus do not claim to have detected a distinct, outer halo in this 
system\footnote{The lack of a need for an outer component is not unprecedented. 
Gonzalez et al.\ (2000) present the light-profile for the BCG+halo in 
Abell 1651, and although it extends to an impressive 670 $h^{-1}$ kpc, 
there is no sign of a transition radius and they show it is remarkably 
well fit by a single $R^{1/4}$ model.}, which may therefore not be a cD galaxy. 
That is, this system may simply have been mis-classified
as a cD galaxy because of excess flux at large radii compared to an $R^{1/4}$
model. 
\\
\\

\noindent
{\em UGC 9799}:
\\

\noindent
Figure \ref{ugc9799}b shows the double S\'ersic fit to this galaxy. The inner
and outer components are both modelled with $n\sim 1$, suggesting that this 
galaxy's surface brightness profile may be well characterized by
a double exponential. Keeping the outer component modelled
with $n=1$ (i.e. exponential) and the inner component modelled with a free
value of $n$ does not significantly change the 
residuals (Figure \ref{ugc9799}c) and shows that
the inner component still has a S\'ersic index $n\sim 1$.
Not surprisingly, the affect of fitting this galaxy with a
double exponential (not shown) does not change the residuals significantly. 
It is unusual though to see such a large elliptical galaxy with
a low value for $n$, since luminous early-type galaxies usually have
high S\'ersic indices (Caon, Capaccioli \& D'Onofrio 1993; 
Graham et al.\ 1996). 

%   R_e is $\sim$40 kpc for the inner component. so the following removed: 
% Is it possible that this galaxy
% has simply been mis-classified, due to poor ground-based 
% resolution. We note that S0
% galaxies can have nearly exponential bulge profiles
% (Aguerri et al.\ 2005b; Balcells et al.\ 2003). 
% SB0 galaxies have been misclassified
% as other classes of object in the past (e.g.\ Aguerri et al.\ 2005b;
% Barth 2007) and so
% this is a question that is worth following.

Figure \ref{struct} shows a series of $I$-band ({\tt F814W}) 
{\tt HST/WFPC2} images of UGC 9799 (taken from the {\tt HST} archive 
and observed as part of program number SNAP-8683 PI: van der Marel).
Images are shown with different stretches, highlighting both low-surface
brightness features (left), and the fact that this galaxy has
an active nucleus (right), which is classified as Seyfert 2 in the NASA
Extragalactic Database (NED). 
The image in the bottom left of Figure \ref{struct} is a type of residual
image, created by fitting the smooth galactic 
starlight with elliptical isophotes and subtracting the original
image from this model. Fine structure emerges using
this technique. In this case low surface brightness features
to the north-east of the nucleus have been revealed.
The image in the bottom right of Figure \ref{struct} is a structure map
(of the kind presented by Pogge \& Martini 2002) and again highlights the
AGN.

For our ground-based surface brightness profiles, the inner
few points (determined from the size of the relevant seeing disk)
are ignored when performing the analytical fits. 
As a result, the bulk of the AGN
contribution will also be ignored, as this will be the same
size as the seeing disk. The fine structure seen in Figure 7 is also
common in central cluster galaxies and can be interpreted as
features, which appear as a result of the merger
processes involved in the formation of cD galaxies. 
From a  morphological point of view, 
UGC 9799 shows nothing that would
not be expected for a cD galaxy, apart from an inner exponential fit. 
Although, this was also observed in NGC 4874 and thus may be more common in
cD galaxies than previously thought.

\section{Discussion}

Constraining the 
surface brightness profiles of the faint envelopes in cD galaxies 
is important for constraining current models of cD galaxy
formation and cluster dynamics.
There is evidence from the globular cluster population, and the 
near-infrared galaxy luminosity function, to suggest that BCGs experienced 
their mergers long ago (Jordan et al.\ 2004; Ellis \& Jones 2004), yet the 
presence of (un-erased) tidal streams (e.g.\ Gregg \& West 1998; Trentham 
\& Mobasher 1998) would appear to favour a more recent formation epoch.
One of the earliest studies of the low surface brightness haloes of
cD galaxies was performed by Carter (1977), who found that the total luminosity
did not converge even at a radius of 300 kpc. This was later
confirmed by Lugger (1984), who additionally
found that a de Vaucouleurs $R^{1/4}$ model
consistently underestimated the surface brightness profiles of cD galaxies
at large radii (see also Schombert 1986). 
These observations have been confirmed in recent years
by deep imaging of central cluster galaxies (e.g.\ Feldmeier et al.\ 2002, 
2004a, b; Lin \& Mohr 2004; Adami et al.\ 2005;
Gonzalez, Zabludoff \& Zaritsky 2005;
Kemp et al.\ 2005; Krick, Bernstein \& Pimbblet 
2006; Liu et al.\ 2005; Mihos et al.\ 2005; 
Zibetti et al.\ 2005) which clearly indicate
the presence of an extended stellar envelope, albeit 
relative to an inner $R^{1/4}$
component. The need for two S\'ersic components, rather than one, is clearly 
illustrated in Gonzalez, Zabludoff \& Zaritsky (2003, their Figs.~1 \& 
2).
Furthermore, it has also been shown that the Petrosian properties
of central cluster galaxies display 
distinct properties, which can be interpreted as an indicator of cD galaxy haloes,
independent of an assumed $R^{1/4}$ light-profile for the central galaxy
(Brough et al.\ 2005; Patel et al.\ 2006).

We note that our photometry alone does not indicate whether the outer 
`component' is physically distinguished from the inner `component'.  In 
general, however, a {\it single} S\'ersic function provides a good fit to 
ordinary elliptical galaxies, with little or no structure in their 
resultant residual profiles. The structure seen in the residual profiles 
for 4 of our 5 cD galaxies is reminiscent of that seen when fitting a 
single S\'ersic function to a spiral galaxy, and is suggestive of two 
distinct components. The failure of a single S\'ersic function to match 
the observed stellar distribution is physical evidence that four of our 
objects are different from ordinary elliptical galaxies, but we caution 
that they may still be single physical entities.

A small fraction of the ICL, and thus envelopes around cD 
galaxies, probably originates from stars 
which have been gravitationally ejected by supermassive black hole binaries 
at the centres of elliptical galaxies within the cluster
(e.g.\ Holly-Bockelmann et al.\ 2006).  Recently, 
Graham (2004) has shown that the central stellar mass deficit in ``core'' 
galaxies --- thought to have formed in ``dry'' mergers --- is roughly 0.1
per cent of their total stellar mass (see also Ferrarese et al.\ 2006).  
This mass deficit is roughly equal to the (combined) 
mass of the central black hole in ellipticals, and is also consistent with 
theoretical predictions on the orbital decay of binary black holes (Ebisuzaki, 
Makino \& Okumura 1991; Milosavljev\'ic \& Merritt 2001). 
Most recently, in 
high-precision, $N$-body simulations, Merritt (2006) has shown that virtually 
all of the mass deficit is generated during the rapid, initial phases of 
binary 
formation, not after the binary becomes hard.  He obtains mass deficits on the 
order of the mass of the binary's larger black hole, and so one can expect the 
intracluster light (from this mechanism) to roughly equal $\sim 0.1$ per
cent of the cluster light in 
spheroids\footnote{This is an upper estimate because in gas rich mergers, gas 
facilitates the decay of the binary black hole, and consequently less stars 
are ejected.}.

Observations of multiple nuclei within  brightest cluster 
galaxies (e.g. Hoessel \& Schneider 1985; Postman \& Lauer 1995;
Seigar, Lynam \& Chorney 2003) and low surface brightness
tidal features (e.g. van Dokkum 2005) are 
considered strong evidence that massive galaxies are growing at the
centres of rich clusters by accreting their less massive neighbours (Hausman
\& Ostriker 1978). Such merger events are
also thought to be partly responsible for the formation of 
extended (low surface brightness) envelopes and intracluster light
(Ostriker \& Tremaine 1975; Ostriker \& Hausman 
1977; Hill \& Oegerle 1998;
Moore et al.\ 1996; Muccione \& Ciotti 2004; Willman et al.\ 2004). 
This is demonstrated in recent semi-analytical models (e.g., Purcell, Bullock 
\& Zentner 2007)
and N-body simulations (e.g., Conroy, Wechsler \& Kravtsov 2007). 
Moreover, close galaxy-galaxy encounters can strip stars from deep within 
their potential well.  These stars may then be liberated by the overall 
cluster tidal field (Merritt 1984; Moore et al.\ 1996), to become what is 
known as the intracluster light (ICL: Zwicky 1951; Welch \& Sastry 1971; 
Oemler 1973; Thuan \& Kormendy 1977). Furthermore, the large numbers of 
ultra compact dwarf galaxies found in galaxy clusters suggests that this 
process may be rather efficient (e.g., Bekki et al.\ 2003; C\^ot\'e 2005; 
Drinkwater et al.\ 2005; Gnedin 2003; Mieske, Hilker \& Infante 2005).  
It is also 
likely responsible for the existence of intracluster: planetary nebulae 
(Arnaboldi et al.\ 2004; Aguerri et al.\ 2005a; Feldmeier et al.\ 2004b; 
Gerhard et al.\ 2005); red giant stars (Ferguson, Tanvir \& von Hippel 
1998; Durrell et al.\ 2002); novae (Neill, Shara \& Oegerle 2005); and 
supernova (Gal-Yam et al.\ 2003).

Our results reveal, given the validity of these processes, that they result
(in three instances) in an exponential-like distribution of stars around the 
central dominant galaxy
More precisely, for three of our five galaxies, an 
inner S\'ersic model plus an outer exponential model provides a good fit 
to the data.  In one galaxy, NGC 6173, no outer exponential model is 
required, and in NGC 4874, the outer light profile is best described with 
an $R^{1/4}$ model rather than an exponential model.

If the envelopes associated with cD galaxies trace a surrounding dark 
matter halo, then one might expect them to be described by a S\'ersic 
function with $n$ around 2.5 to 3. This is because hierarchical 
$\Lambda$CDM simulations produce a near universal profile shape for dark 
matter halos on all scales, the projection of which is well described by a 
S\'ersic $R^{1/n}$ model with $n\sim 3$ ({\L}okas \& Mamon 2001; Merritt 
et al.\ 2005, 2006).

Curiously, $N$-body simulations of cold collapses (and disk galaxy
mergers) also result in haloes (and merger remnants) 
having, in projection, $R^{1/3}$-like profiles (e.g.,
Willman et al.\ 2004, their Fig.7; Aceves, Velaquez \& Cruz
2006; Merritt et al.\ 2006). Interestingly, a
closer inspection of old data (e.g.\ Figures 4--6 in van Albada 1982) 
reveals obvious and systematic deviations from the $R^{1/4}$ model 
in the sense that an $R^{1/n}$ model with $n<4$ provides a better fit to
the data presented there. 
%
% $N$-body simulations of cluster-sized
% dark matter haloes are known to have density profiles similar to
% the Navarro, Frenk \& White (1996; hereafter NFW) profile.
% These profiles have recently been shown to be well described
% with either the Einasto (1965) or Prugniel-Simien (1997) model (e.g.\ Merritt
% et al.\ 2006), the projection of which
% yields S\'ersic profiles with index $n$
% ranging from 2.2 to 3.5 (Merritt et al.\ 2006; Dalcanton \&
% Hogan 2001; {\L}okas \& Mamon 2001).  
% 
If the halo or intracluster light around cD galaxies traces the
dominant dark matter potential, one does not
expect this envelope to be described with an $R^{1/4}$ model.
Of course, baryonic processes may well result in a different
stellar distribution to the dark matter distribution,
and studies suggest this can lead to a flattening of the inner
density profile (Nipoti et al.\ 2004), which is effectively
equivalent to a reduction in the S\'ersic index.

\begin{table*}
\caption{Results using the double S\'ersic model to derive ratios of physical
  parameters. Column 1: galaxy name. Column 2: effective radius of the central
  galaxy. Column 3: effective radius of the ICL or envelope. Column 4:
  galaxy-to-envelope size ratio, given by the ratio of the effective radii
  $R_{\rm e, 1}/R_{\rm e, 2}$. Column 5 and 6: envelope-to-total ratio 
calculated by extrapolating the profiles to infinity, $(E/T)_{\rm tot}$, 
and 300 kpc, $(E/T)_{300}$.}
\begin{tabular}{llllll}
\hline
Galaxy         & $R_{\rm e, 1}$     & $R_{\rm e, 2}$     & $R_{\rm e, 1}/R_{\rm e, 2}$ & $(E/T)_{\rm tot}$ & $(E/T)_{300}$ \\
               & (kpc)          & (kpc) \\
    1          &    2           &        3        &        4          &     5    &      6    \\
\hline
GIN 478        & 19.9$\pm$1.6   & 261.2$\pm$84.9 & 0.076$\pm$0.019     & 0.82$\pm$0.13     & 0.73$\pm$0.13 \\
NGC 3551       & 53.3$\pm$11.6  & 129.7$\pm$30.9 & 0.411$\pm$0.072     & 0.59$\pm$0.04     & 0.47$\pm$0.05 \\
NGC 4874       & 1.8$\pm$0.1    & 108.1$\pm$12.3 & 0.017$\pm$0.002     & 0.98$\pm$0.01     & 0.95$\pm$0.01 \\
NGC 6173       & 1073.1$\pm$296.8 & -- & -- & -- & -- \\
UGC 9799       & 10.5$\pm$2.0   & 48.4$\pm$9.0   & 0.217$\pm$0.042     & 0.77$\pm$0.14     & 0.76$\pm$0.14 \\
\hline
\end{tabular}
\end{table*}

Demarco et al.\ (2003) have analysed the distribution of X-ray
gas in a sample of 24 real galaxy clusters.  They found it was
well described by a S\'ersic function having values
$0.8 < n < 2.3$.   With the exception of NGC 4874, this range
is in good agreement with the values reported here for the
outer component of our cD galaxies, and suggests that this
envelope is indeed tracing the (azimuthally-averaged)
intracluster light.  Moreover, our exponential-like
outer profiles match the exponential ICL profile in Abell 3888
(Krick et al.\ 2006).  When more observations become available,
it will be interesting to see whether the distribution of the
intracluster stellar probes, such as planetary nebula and globular clusters, follow
an exponential or an $R^{1/4}$ radial distribution. It will
also be interesting to know if the intragroup light (e.g.,
Da Rocha \& de Oliveira 2005; Faltenbacher \& Mathews 2005;
White et al.\ 2003) behaves in a similar or different manner.

The mean BCG+ICL light-profile obtained from the stacked cluster image 
reported in Zibetti et al.\ (2005) is plotted using a {\it linear} radial 
axis in Zibetti \& White (2004, their Fig.~1)\footnote{Zibetti \& White 
(2004) use 654 clusters, while Zibetti et al.\ (2005) use 683.}.  One can 
immediately see by eyeball examination
that the outer light-profile is  well approximated
by an exponential (i.e., a straight line in that figure).  This is in good 
agreement with our independent data and the (cluster halo) X-ray data from Demarco et 
al.\ (2003), but at odds with the $R^{1/4}$ model used in Gonzalez et al.\ 
(2005), and at odds with a projected NFW model. 
We do however note that the shape of the ICL profile in Zibetti et al.\
(2005) is different, suggestive of a S\'ersic index greater than 1.
This difference arose from their new corrections for mask incompleteness
and their new method of determining the sky background. The latter involved
simultaneously fitting an NFW model for the ICL and some constant value
for the sky background level. The problem with such an approach is that the 
fitted constant effectively modifies the real ICL profile to produce (as
best as it can) an NFW profile, even when the real ICL may not have such a 
form. Fitting an $R^{1/4}$ model from 150 to 500 kpc, Zibetti et al.\ (2005,
their section 5.1) report an effective radius $R_{\rm e}$ of 250--300 kpc for the
ICL, somewhat larger than the values we observe in our sample (Table 4).

\subsection{Relative contribution of the stellar envelope to the total luminosity of cD galaxies}

We use the analytical fits from Table~\ref{Tab2} to determine the relative 
contribution of the stellar envelope to the total luminosity of the galaxy
plus envelope, i.e.\ the 
envelope-to-total ratio. We compute two estimates of this flux ratio. One of 
these, $(E/T)_{300}$, comes from truncating the models at a radius of 300 
kpc, as done in Gonzalez et al.\ (2005). 
% This appears later: 
% who report typical $(E/T)_{300}$ 
% ratios around 0.9 but as low as 0.4 using an $R^{1/4} + R^{1/4}$ 
% parameterisation. Zibetti et al.\ 
% (2005) use a truncation radius of 500 kpc, and report an $E/T$ ratio of 33 
% per cent, obtained from $(10.9\pm5.0)/((10.9\pm5.0)+(21.9\pm3.0))$. 
Our second estimate assumes no 
truncation, and is denoted $(E/T)_{\rm tot}$. Due to the extended nature 
of the envelope, our $(E/T)_{\rm tot}$ values are somewhat different
when compared to our 
$(E/T)_{300}$ values, although only by $\sim$10 per cent at most. 
Both quantities are listed in Table 4. It should be noted that since the
radial extent of our surface brightness profiles is $\sim$100 kpc, although 
223 kpc for GIN~478, that we
have to extrapolate to estimate both the $(E/T)_{300}$ and $(E/T)_{\rm tot}$
values. 

In most cases the extended stellar envelope contributes around 60
to 80 per cent of the total $R$-band luminosity when no truncation
radius is applied and the models are extrapolated to infinity.
When the profiles are truncated at 300 kpc, the envelope
contributes between $\sim$45 to 80 per cent. The exception is
NGC 4874 (the only galaxy with an $R^{1/4}$ envelope) which has
$E/T\simeq$ 95--98 per cent. 
% depending on whether a truncation radius is applied (the smaller value) 
% or extrapolation is applied (the larger value). 
These flux ratios lie in the same range found by
Gonzalez et al.\ (2005), who reported $E/T$
ratios of around 0.9 but as low as 0.4 (their Figure 7).
However, Zibetti et al.\ (2005) find a slightly lower value. 
They find that the ICL contributes 10.9 percent
to the total cluster light and the central galaxy contributes 21.9 per cent. 
This is equivalent to an $E/T$ ratio of $\sim$33 per cent.
However, taking into account the errors in the measurements, $E/T$ ratios
in the range 25 to 45 per cent are allowed. The high end of their range is
therefore consistent with the results found here.

For the three cDs best described with an $R^{1/n}$ galaxy plus exponential
envelope, the galaxy-to-envelope size ratio (given by the ratio of
the effective radii $R_{\rm e, 1}/R_{\rm e, 2}$) ranges from $\sim$0.1
to $\sim$0.4.  In contrast, using double $R^{1/4}$ models, Gonzalez et al.\ (2005) report ratios of $\sim$0.1 down
to $\sim$0.025, i.e.\ envelopes 10 to 40 times larger in size
than the inner component.  One of our remaining two galaxies
appears to have no distinct envelope, and the other has an
$R^{1/4}$ envelope 60 times greater in size than the
central galaxy.

\section{Summary}

We have observed 5 cD galaxies to a depth of $\mu_R=26.5$ mag arcsec$^{-2}$,
and we have determined the shapes of the surface brightness profile of their 
outer stellar envelopes. 

The results of previous attempts to model the intracluster light or extended 
stellar envelope suggested that a universal model applies. 
For example, Gonzalez et al.\
(2005) report that both the central part of the cD and
the intracluster light are both well described by an $R^{1/4}$
surface brightness model.
In general, previous studies of this kind have only tried fitting a surface
brightness model of one kind. Our approach of fitting a S\'ersic model to the
extended halo provides a means to actually measure, rather than 
pre-ordain the actual stellar distribution, albeit within the confines
of the S\'ersic model.

Our analysis suggests that the surface brightness profiles of cD galaxies
(including their envelopes) are best modelled by a double S\'ersic
function. While an inner $R^{1/4}$ model is sufficient for some cDs, we
have found that the inner S\'ersic index can vary 
significantly from object to object (from $n\sim 1$ to $n\sim 7$).
An outer exponential model seems appropriate for three of our four, 
2-component systems. 
% (see also Krick et al.\ 2006),
One galaxy (NGC 4874) appears to have an $R^{1/4}$ envelope. 
%%    removed to appease the referee Zibetti: 
% It is interesting to note that an eyeball examination of SDSS stacked
% clusters, on average, reveals that the ICL has a light profile that is close
% to exponential (Zibetti et al.\ 2005).
% 
%%   Too much repetitive detail, rather than a summary.
% One galaxy (UGC 9799) seems to be best fit with
% an inner exponential law and an outer exponential law. This is unusual
% for such a luminous early-type galaxy. 
% However, on inspection of a {\tt WFPC2/HST} image,
% no unexpected morphology is seen and no explanation can be given for the low
% S\'ersic index for this galaxy's inner component. 
% One other galaxy (NGC 4874) is also best fit with an inner exponential
% component and an $R^{1/4}$ envelope.
% Another galaxy, NGC 6173, has a surface brightness profile which
% is consistent with a single component, and may simply have been mis-classified
% as a cD galaxy, because of excess flux at large radii compared to an $R^{1/4}$
% model. The other two galaxies have surface brightness profiles
% that can be described with an inner S\'ersic fit (with $n \sim 4$)
% and an outer diffuse 
% component with an exponential surface brightness profile.
Typically, when present, the envelope contributes  
around 60 to 90 per cent of the total (galaxy + ICL) light, 
and the galaxy-to-envelope size ratio is $\sim$0.1 to $\sim$0.4. 
% which is consistent with the studies of
% Gonzalez et al.\ (2005) but somewhat higher than the 33 per cent value
% reported in Zibetti et al.\ (2005). 

\section*{Acknowledgments}
The Jacobus Kapteyn Telescope (JKT) was operated on the island of La Palma by 
the Isaac Newton Group in the Spanish Observatorio del Roque de los Muchachos 
of the Instituto de Astrofisica de Canarias. This research has made use of the 
NASA/ IPAC Extragalactic Database (NED) which is operated by the Jet 
Propulsion Laboratory, California Institute of Technology, under contract with 
the National Aeronautics and Space Administration. This work made use of
observations made with the NASA/ESA Hubble Space Telescope, obtained from
the data archive at the Space Telescope Science Institute. STScI is operated
by the Association of Universities for Research in Astronomy, Inc.\ under 
NASA contract NAS 5-26555. 
MSS acknowledges partial support from a Gary McCue Fellowship through the 
Center for Cosmology at UC Irvine.
The authors wish to thank Paul Lynam for useful
comments and suggestions. 
The authors also would like to thank the 
referee, Dr.\ S.\ Zibetti, for useful suggestions and comments.

\label{lastpage}

\end{document}